# Sequential Edge-Epitaxy in 2D Lateral Heterostructures

Prasana K. Sahoo[1]*, Shahriar Memaran[2], Yan Xin[2], Luis Balicas[2] and Humberto R. Gutiérrez[1]*

[1] *Department of Physics, University of South Florida, Tampa, Florida 33620, USA*

[2] *National High Magnetic Field Laboratory, Florida State University, Tallahassee, FL 32310, USA*

* *e-mail:* humberto3@usf.edu; psahoo@usf.edu

Abstract

Two dimensional (2D) heterojunctions display a remarkable potential for application in high performance, low power electro-optical systems.[1] In particular, the stacking of transition metal dichalcogenides (TMDs) can open unforeseen possibilities given their tunability, material and thickness dependent band gaps, and strong light−matter interaction.[2] p−n junctions based on vertically stacked heterostructures have shown very promising performance as tunneling transistors,[3,4] light emitting devices and photodetectors,[5-7] and as photovoltaic cells .[8,9] Although complex vertical heterostructures[5,7] were fabricated *via* van der Waals stacking of different 2D materials,[4] atomically sharp multi-junctions in lateral heterostructures is a quite challenging task, but a viable route towards the development of commercial applications. Previously reported approaches to obtain single-junction lateral heterostructures of the type $MoX_2$-$WX_2$ (X= S and/or Se), involve either a single-step[10-12] or a two-step growth process.[13,14] However, these methods lack the flexibility to control the lateral width of the TMD domain as well as its composition. Here, we report a simple and yet scalable synthesis approach for the fabrication of lateral multi-junction heterostructures based on the combination of different TMD monolayers [$MoX_2$-$WX_2$ ($X_2$ = $S_2$, $Se_2$ or SSe)]. Atomically sharp lateral junctions are sequentially synthesized from solid precursors by changing only the reactive gas environment in the presence of water vapor. This allows to selectively control the water-induced oxidation[15], volatilization[16] and hence the relative amount of a specific metal oxide vapor, leading to the selective edge-epitaxial growth of either $MoX_2$ or $WX_2$. Spatially dependent photoluminescence and atomic-resolution images confirm the high crystallinity of the monolayers and the seamless lateral connectivity between the different TMD domains. These findings could be extended to other families of 2D materials, and creates the foundation towards the development of complex and atomically thin in-plane super-lattices, devices and integrated circuits.[17]

A key step for the fabrication of TMD-based heterostructures is to control the relative amount of precursors in the gaseous phase from the solid sources during a one-pot synthesis strategy. In general, compounds based on $MX_2$ (where: M=[W, Mo] and X= [S, Se]) have high melting points. Therefore, they are not a common choice as precursors for vapor transport synthesis, although dissociation has been observed by mass spectroscopy at temperatures ranging between 930 and 1090 °C.[18] The presence of water vapor can significantly influence their physicochemical properties, such as oxidation and the formation of volatile species.[15, 16] The enhanced volatility of Mo/W, and of their known oxides above 1000 °C, in the presence of water vapor, has already been reported in the early 1950s.[16, 19, 20] Reaction of $MoS_2$ and water vapor was initially reported by Cannon et al.[15] which indicates that $MoS_2$ undergoes quick oxidation above 450 °C. However, a minimum temperature of 1000 °C is required to generate an equilibrium concentration of $H_2S$ in the reactive environment relative to the concentration of $H_2O$.[21] In this report, $MoX_2$-$WX_2$ lateral heterostructures were grown using a solid source composed of $MoX_2$ and $WX_2$ powders placed side-by-side within the same boat at high temperatures. The selective growth of each material was controlled independently only by switching the carrier gas, i.e. from $N_2$+$H_2O$ (v) to Ar+$H_2$ (5%). $N_2$+$H_2O$ (v) promotes the growth of $MoX_2$. Switching to Ar+$H_2$(5%) stops the growth of $MoX_2$ and promotes the growth of $WX_2$. Figure 1a and 1b (left panels) show optical images of a lateral heterostructure containing a core composed of monolayer $MoSe_2$ (darker contrast) and a $WSe_2$ monolayer shell (lighter), both grown on $SiO_2$/Si substrate, containing different shell lateral sizes (Extended data Fig. 1a-1h). The sizes of both the core and the shell were controlled by choosing the growth time of each individual section. These $MoSe_2$-$WSe_2$ single-junction monolayers predominantly display, an equilateral triangle geometry. Noticeably, the nucleation of the consecutive material ($WSe_2$) happens mainly at the edges of the $MoSe_2$ triangular monolayers resulting in a lateral epitaxial growth, as shown below by the Scanning Transmission Electron Microscopy (STEM) analysis.

In general, the typical sample area is at least 5 x 5 mm$^2$ but hundreds of similar monolayer lateral heterostructures can be found on the same substrate. The average size of the heterostructure islands varies with their position on the substrate due to the temperature profile of the furnace (Fig. 1l); and hence is a function of the substrate temperature for a particular growth condition (Extended data Fig. 1i). The most important outcome of our method is the ability to sequentially grow multiple lateral hetero-junctions in a single monolayer island by merely switching back and forth the carrier gas, under optimized conditions. This new approach simplifies considerably the synthesis of periodic lateral heterostructures composed of TMDs or of any other layered materials. Figure 1c shows an optical image of three-junction lateral heterostructures consisting of alternate monolayers of $MoSe_2$ (dark) and $WSe_2$ (light) [i.e. $MoSe_2$-$WSe_2$-$MoSe_2$-$WSe_2$] with thicknesses of 42-11-10-11 μm, obtained under growth times of 720-60-240-60 s, respectively. Figure 1d shows a magnified image of the area enclosed by the white dashed box in Fig.1c. The outer $WSe_2$ edge extends up to a length of 285 μm, which is the longest ever reported value on any seamless monolayer heterostructure.[14] By increasing the number of gas-switching cycles, one can synthesize more complex periodic lateral heterojunctions, as shown in the Fig. 1e-g.

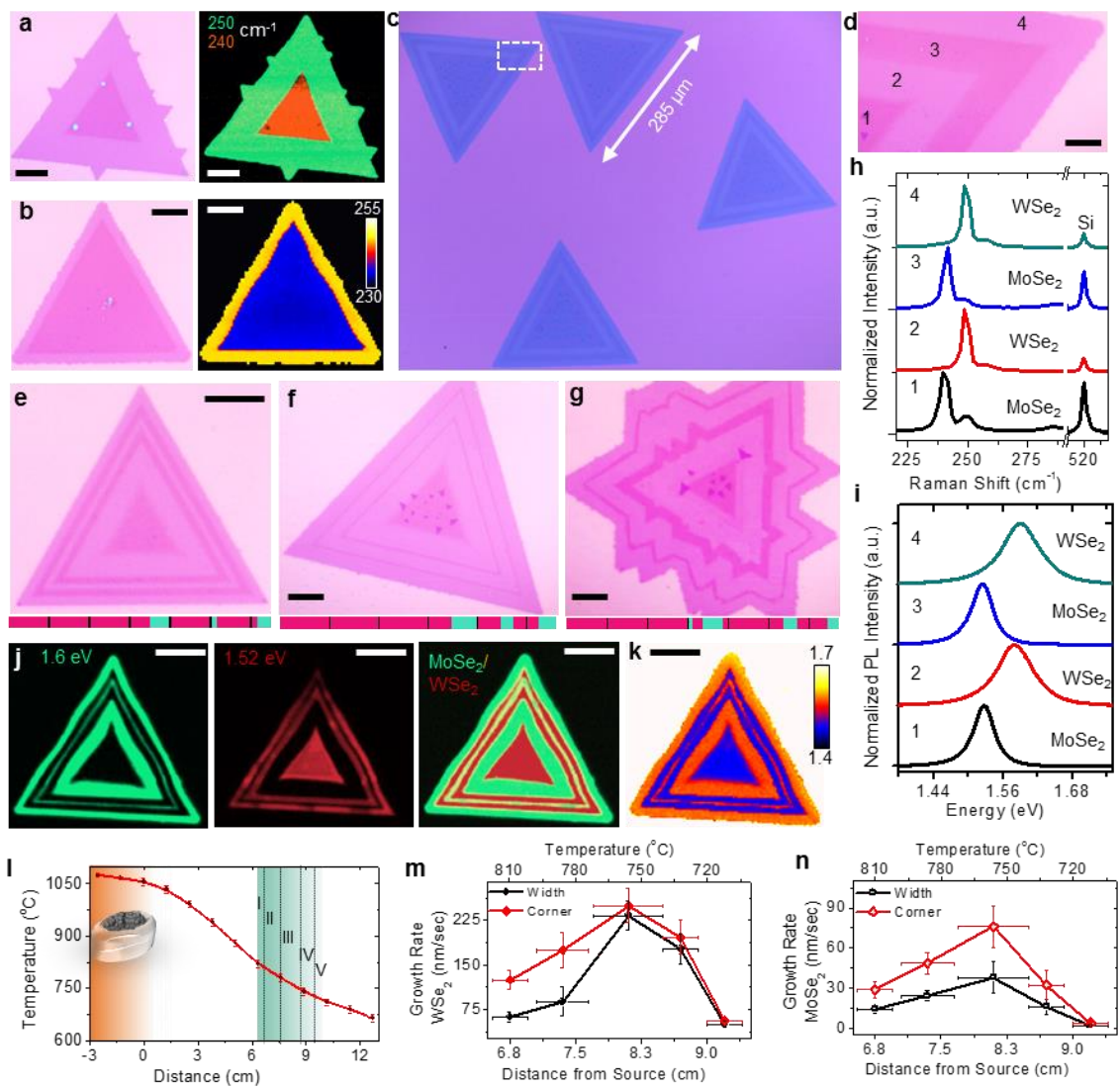

**Figure 1. Growth of MoSe$_2$-WSe$_2$ lateral heterostructures. a, b,** Optical images of MoSe$_2$-WSe$_2$ single-junction lateral heterostructures with different WSe$_2$ lateral widths. Right panel **a**, Composite Raman intensity maps for 250 cm$^{-1}$ (green) and 240 cm$^{-1}$ (red). Right panel **b**, Raman peak position map. **c,** Low magnification optical image of four island consisting of three-junction heterostructures. **d,** Larger magnification optical image of the dashed square box in (c). **h,** and **i,** Raman and photoluminescence (PL) spectra, respectively, at positions 1, 2, 3 and 4 in (d). **e, f,** Optical images of five-junction heterostructures with different widths for the MoSe$_2$ stripes. **g,** Seven-junction heterostructure with variable widths for both the MoSe$_2$ and the WSe$_2$ domains. The underlying color bar (in **e**,**f**,**g**) depicts the growth time scale, i.e. from left to right (pink – MoSe$_2$, green – WSe$_2$)- each division (black lines) corresponds to ~ 120 s. **j, k,** PL intensity and peak position mapping, respectively, for the heterostructure in (**e**). **l,** Furnace temperature profile, regions I, II, III, IV and V correlates with the data points in (m and n). **m, n,** Growth rates along different directions for WSe$_2$ and MoSe$_2$ islands, respectively. Scale bars, 10 μm.

The five-junction lateral heterostructures shown in Figs. 1e and 1f, differ in the width of the MoSe$_2$ stripes; this was achieved simply by reducing the growth time of the MoSe$_2$ stripes from 120 sec (Fig. 1e) to 60 sec (Fig. 1f), Figure 1g shows a seven-junctions heterostructure with controlled variable lateral width for each section of a given material. In this case, the lateral growth remains conformal even when new "facets" emerge at the different stages of the growth. Figure 1m illustrates the growth rates for $MoSe_2$ and $WSe_2$ monolayers, respectively, as a function of the growth temperature (or different substrate positions within the temperature profile of the furnace (Fig. 1l). It indicates that the growth rate for $WSe_2$ is two-to-four times larger than the one for the $MoSe_2$ monolayer.

Raman and micro-photoluminescence spectroscopies were used to probe the local optical properties as well as the spatial chemical distribution of the heterostructures. Figure 1h shows Raman spectra collected from different points on the heterostructure (marked as 1, 2, 3 and 4 in Fig. 1d). We detect two dominant peaks at 240 cm$^{-1}$ (points 1,3) and 250 cm$^{-1}$ (points 2,4) which are characteristic of monolayer $MoSe_2$ and $WSe_2$, respectively.[11] The peak at 240 cm$^{-1}$ corresponds to the $A_{1g}$ phonon mode of monolayer $MoSe_2$, whereas the additional shoulder at 249 cm$^{-1}$ (points 1,3) has been assigned to the $E_{2g}^2$ shear mode of $MoSe_2$ at the M point of the Brillouin zone (Extended data Fig. E2a-c).[22] In contrast, in regions 2 and 4 the highest peak is observed at 250 cm$^{-1}$, which matches the $A_{1g}$ mode of monolayer $WSe_2$. The peak at 260 cm$^{-1}$ is attributable to the second-order Raman mode of the $LA(M)$ (130 cm$^{-1}$) phonons. The spatial distribution of the $MoSe_2$ and $WSe_2$ domains was further investigated through Raman intensity maps at 240 cm$^{-1}$ and 250 cm$^{-1}$, respectively (Fig. 1a, right panel). The triangular outer shell (light pink) exhibits a uniform distribution of the $WSe_2$ Raman signal while the triangular core exhibits the characteristic $MoSe_2$ Raman signal (Fig. 1a). The Raman peak position map (Fig. 1b, right panel) also corroborates this chemical distribution (detailed in Extended data Fig. E2c-2l).

The PL spectra (Fig. 1i) acquired from points 1-4 correlate well with the chemical distribution obtained by the Raman spectroscopy. At positions 1 and 3, only a strong peak is observed around 1.52 eV associated to the emission from a direct exciton in monolayer $MoSe_2$. Whereas at points 2 and 4 the PL peak is centered around 1.6 eV which corresponds to the excitonic transition in monolayer $WSe_2$.[11] The integrated PL intensity (Fig. 1j) and the peak position maps (Fig. 1g) of the heterostructure in Fig 1e, confirm the alternate formation of concentric triangular domains of $MoSe_2$ and $WSe_2$ monolayers (individual maps in Extended data Figs. E3c-e). Figure 2c shows the normalized PL spectra at different positions along a line spanning three-junctions within the heterostructure (see inset), i.e. from the $MoSe_2$ core (top) towards the outer $WSe_2$ shell. The contour plot of the normalized PL intensity (Fig. 2d) as a function of the position across the heterostructure, clearly shows the evolution of the excitonic transition from one material to the other. Across the 1$^{st}$ junction (top), the $MoSe_2$ PL peak (at 1.53 eV) gradually shifts to higher energies until it reaches the 1.60 eV value corresponding to the $WSe_2$ domain, for a total shift of 70 meV. On the other hand, at the 2$^{nd}$ and 3$^{rd}$ junctions, there is an abrupt change in the PL peak position; this abrupt evolution of the distinct peaks for $MoSe_2$ (1.53 eV) and $WSe_2$ (1.60 eV) clearly indicates the formation of sharp interfaces between these regions. At the abrupt interfaces, the PL spectrum is characterized by an overlap of both peaks, owing to the 500 nm laser spot size, which simultaneously probe both sides of the interface. Although, junctions 2 and 3 are very sharp, it is worth noticing that junction 3 is slightly smoother than junction 2. We have consistently observed this behavior in all samples. For instance, a contour plot of a five-junction lateral monolayer heterostructure (Fig. 2e) shows similar band alignment with very sharp interfaces (at junctions 2 and 4), while junctions 3 and 5 are slightly smoother. These results indicate that a transition from a $MoSe_2$ to a $WSe_2$ domain results in a slightly diffuse interface, whereas the transition from $WSe_2$ to $MoSe_2$ produces atomically sharp interfaces, as confirmed by atomic resolution STEM images.

The crystallographic structure of the heterostructures and the quality of the junctions were investigated through electron diffraction (ED) and atomic resolution high-angle annular dark-field (HAADF) imaging (Z-contrast) in an aberration-corrected STEM. The thinnest two-dimensional sheets of $MX_2$ display the 2H-phase with the hexagonal $D_{3h}$ point group symmetry. Since in our TEM studies, the electron probe was oriented along the direction perpendicular to the plane of the films, the atomic-resolution Z-contrast images consist of a honeycomb-like atomic arrangement, which is expected for $MoSe_2$ (Fig. 2a) and $WSe_2$ (Fig. 2b) monolayers. Notice that for $MoSe_2$ the intensity of the scattered electrons at the Mo and $Se_2$ sites is very similar. However, in the case of $WSe_2$ the intensity at the W sites is approximately twice that at the chalcogen sites. This allows us to study the chemical composition at the atomic level through a simple inspection of the Z-contrast images. Figures 2a and 2b were taken from different regions of the heterostructures (as indicated by the black arrows) proving that the evaporation-deposition process is very selective responding to the instantaneous growth conditions, even though both solid materials ($MSe_2$ and $WSe_2$) are present in the source precursor.

Figure 2f shows a narrow $MoSe_2$ 2D-ribbon (width ~ 350 nm; optical image in Fig. 1f) in between two $WSe_2$ regions, where the sequence of growth goes from right to left. The atomic-resolution images of both interfaces are shown in Figs. 2g and 2h. The first interface $WSe_2 \rightarrow MoSe_2$ (Fig. 2h) is atomically sharp with very small roughness, while the second interface $MoSe_2 \rightarrow WSe_2$ (Fig. 2g) is more diffuse. Both Fourier transform (FT) spectra indicate single-crystalline structure, suggesting that both interfaces result from an epitaxial lateral growth. Both interfaces are approximately parallel to the zigzag metallic direction. Notice that, since the primitive unit cell is composed by two different atoms, there are two distinct zigzag directions, one containing metal atoms and the other chalcogen atoms. The composition profiles shown at the bottom of Figs. 2g and 2h, represent the atomic fraction of W per vertical atomic column, where each column was taken along the zigzag direction parallel to the interface. From these profiles, average interface roughnesses of 6 nm (21 atomic columns) and of 1 nm (4 atomic columns), were calculated for the diffuse and the sharp interfaces, respectively. These TEM observations are in agreement with the previously shown PL line profile, and support the conclusion that $WSe_2 \rightarrow MoSe_2$ transitions produce atomically abrupt interfaces.

We used the same methodology to control the growth of sequential lateral heterostructure of sulfide ($MoS_2$ and $WS_2$) monolayers (Extended Figure E4). Figure 3a shows the optical image of an alternating three-junction heterostructure ($MoS_2$-$WS_2$-$MoS_2$-$WS_2$), each section was grown (starting with $MoS_2$) for 12-2-4-1.5 minutes, respectively; thereby alternating the carrier gas between $N_2+H_2O$ and $Ar+H_2$ during the synthesis. The corresponding SEM image of its right-top corner is shown in Fig. 3b. Remarkably, the lateral width of an individual layer within a sequential heterostructure can also be modulated in a manner similar to the previously demonstrated case of the $MoSe_2$-$WSe_2$ based heterostructures (more information in Extended Data Fig. E4a-d). However, the morphology of the sulfide-based heterostructures is different from the conformal growth of selenides-based heterostructures. For sulfides, the observed $MoS_2$ cores formed truncate triangles presenting both metal and chalcogen-terminated zigzag edges. The subsequent growth of $WS_2$ happened selectively in one of these edges, leading to $WS_2$ sections with a convex isosceles trapezoid shape. The successive $MoS_2$ growth follows the same behavior.

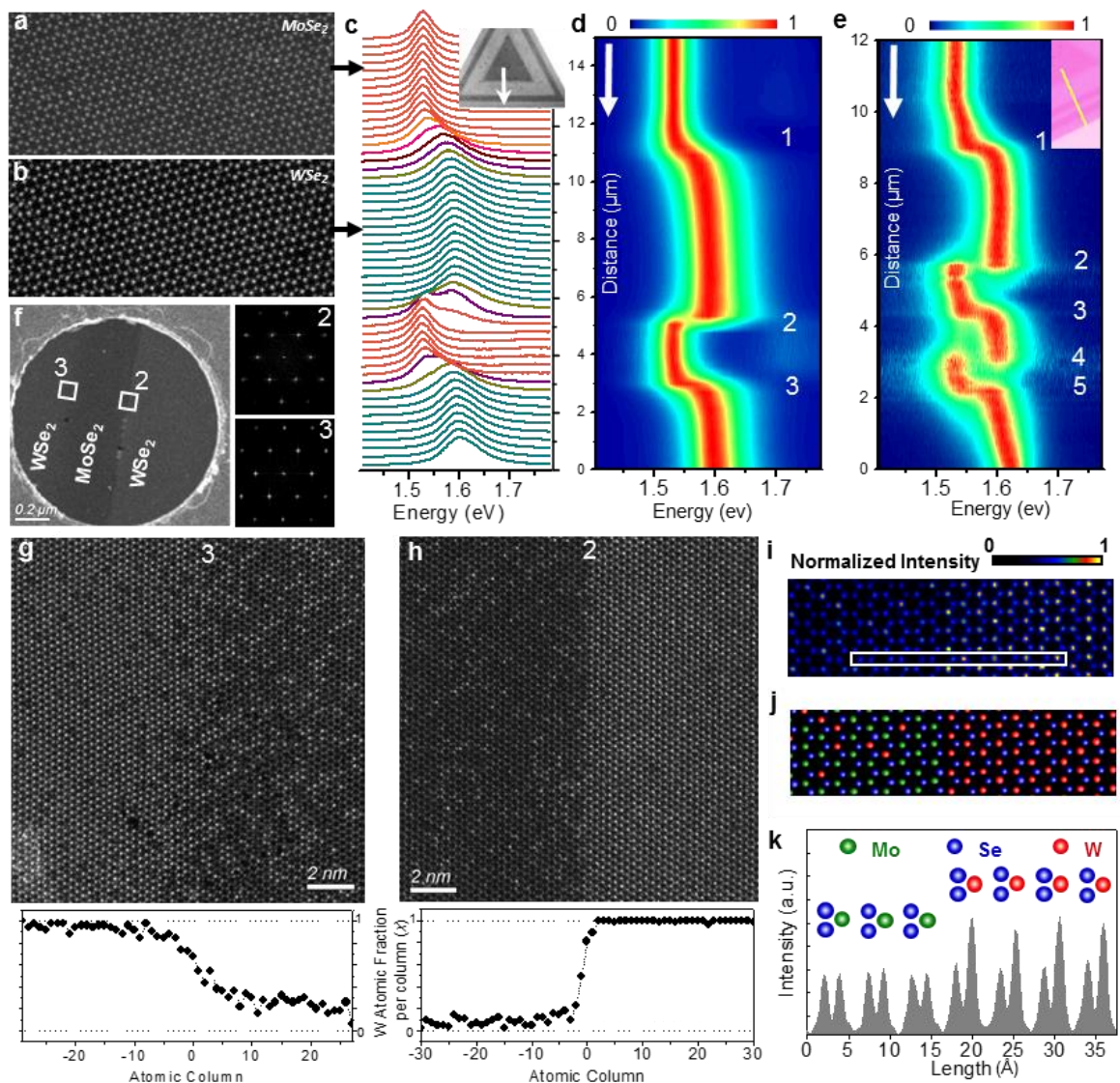

**Figure 2. Multi-junction lateral heterostructures and interfaces based on $MoSe_2$ and $WSe_2$. a, b**, Z-contrast HRTEM images of pure $MoSe_2$ and $WSe_2$, respectively. **c**, Normalized PL spectra from a line scan along the three-junctions, as indicated by the arrow in the inset. **d**, Contour color plots of the normalized PL intensity extracted from the three-junctions in (c). **e**, PL intensity for a five-junction heterostructure (see inset). White arrows indicate the growth direction. **f**, Low magnification Z-contrast TEM image of a 350 nm

wide MoSe$_2$ ribbon. **g**, **h**, Atomic resolution HAADF-STEM images of the left and right interfaces, respectively. Insets are the corresponding FT pattern. **i**, **j**, Scattered electron intensity color plot and associated atomic ball model, respectively, for the region of the junction in (**h**). **k,** Electron intensity profile along the white box in (**i**).

Figure 3c shows the Raman spectra acquired at different positions (i.e. points 1 through 4) of the three-junction lateral heterostructure (Figs. 3a and 3b). The sequential formation of well-defined MoS$_2$ and WS$_2$ domains is confirmed by the distinct spatial location of their corresponding phonon modes. The inner triangle (point 1) exhibits in-plane and out-of-plane vibrational modes at 384 ($E_{2g}$) and 405 cm$^{-1}$ ($A_{1g}$) respectively, which are consistent with monolayer MoS$_2$. At point 3 (MoS$_2$ domain), the $E_{2g}$ mode slightly red shifts by 2 cm$^{-1}$, whereas the $A_{1g}$ mode remains unchanged. This may be attributed to the softening of the in-plane modes as the result of a small alloying in these regions when transitioning from WS$_2$ to MoS$_2$. In contrast, the Raman spectra acquired from points 2, and 4 contain a combination of first-order optical modes ($E^1_{2g}$ at 355 cm$^{-1}$ and $A_{1g}$ at 418 cm$^{-1}$) as well as second-order Raman peaks associated with WS$_2$. The most intense peak at 350 cm$^{-1}$ (*2LA(M)*) is the result of a double resonance Raman process characteristic of monolayer WS$_2$.[23] At the interfaces (1-2, 2-3 and 3-4), the Raman spectra is mostly composed of a superposition of the vibrational modes of the MoS$_2$ and the WS$_2$ domains (Supporting information Table 1). A maximum relative peak distance ($\Delta\omega_{MoS_2}$) of 27, 23 and 26 cm$^{-1}$ was calculated for the junction region denoted as 1-2, 2-3 and 3-4, respectively. However, the Raman peak position corresponding to the WS$_2$ domain did not show any pronounced variation (± 1 cm$^{-1}$). Indeed, the $E_{2g}$ mode is associated with the in-plane vibrations involving the transition metal atoms (Mo / W) and the S atoms, while the out-of-plane $A_{1g}$ mode involves mainly the S atoms. This makes the $E_{2g}$ mode more sensitive to the alloying of the metallic atoms. The observed red shift of the $E_{2g}$ mode in the MoS$_2$ domain, when compared to that of the $E^1_{2g}$ mode of the WS$_2$ domain, indicates the incorporation of a small fraction of W atoms in the MoS$_2$ domain. This mild alloying effect is slightly more pronounced for the 1-2 and 3-4 interfaces where there is a transition from MoS$_2$→WS$_2$. The red shift is smaller for the 2-3 interface (WS$_2$→MoS$_2$), suggesting that WS$_2$→MoS$_2$ transitions produce sharper hetero-junctions. These results are in agreement with the previously discussed observations on the selenide-based heterostructures. The PL spectra in Fig. 3d further agree with the Raman study. The observation of PL peaks associated with direct excitonic transitions around 1.84 eV (MoS$_2$ domains at points 1, 3) and 1.97 eV (WS$_2$ domain at points 2, 4) are consistent with the monolayer nature of the heterostructure. The PL emissions from the respective domains did not show any noticeable variation (< 7 meV), except in the vicinity of the interfaces. The PL spectra collected from the interfaces also exhibit the superposition of two well-resolved peaks corresponding to the simultaneous excitation of MoS$_2$ and WS$_2$ domains at either side of the junctions. The PL peak position associated with the MoS$_2$ domain slightly blue-shifts by 25 meV and by 10 meV at the interfaces 1-2 and 3-4, respectively. No shift was observed at the 2-3 interface implying a very sharp interface. The modulation of the optical bandgap across the heterostructure can be better visualized through a color contour plot of the normalized PL emission collected by scanning along a line perpendicular to the junctions (see Fig. 3e). From this plot, it is evident that the emission energy in the $WS_2$ domains did not change across the line, reflecting the chemical homogeneity within the domain. The PL peak associated with $MoS_2$ remains at a constant energy, while displaying a small blue shift in the vicinity of interfaces 1-2 and 3-4. Notably, for all three interfaces one observes a sharp discontinuity in the position of the PL emission, indicating an efficient modulation of the optical bandgap across the junctions.

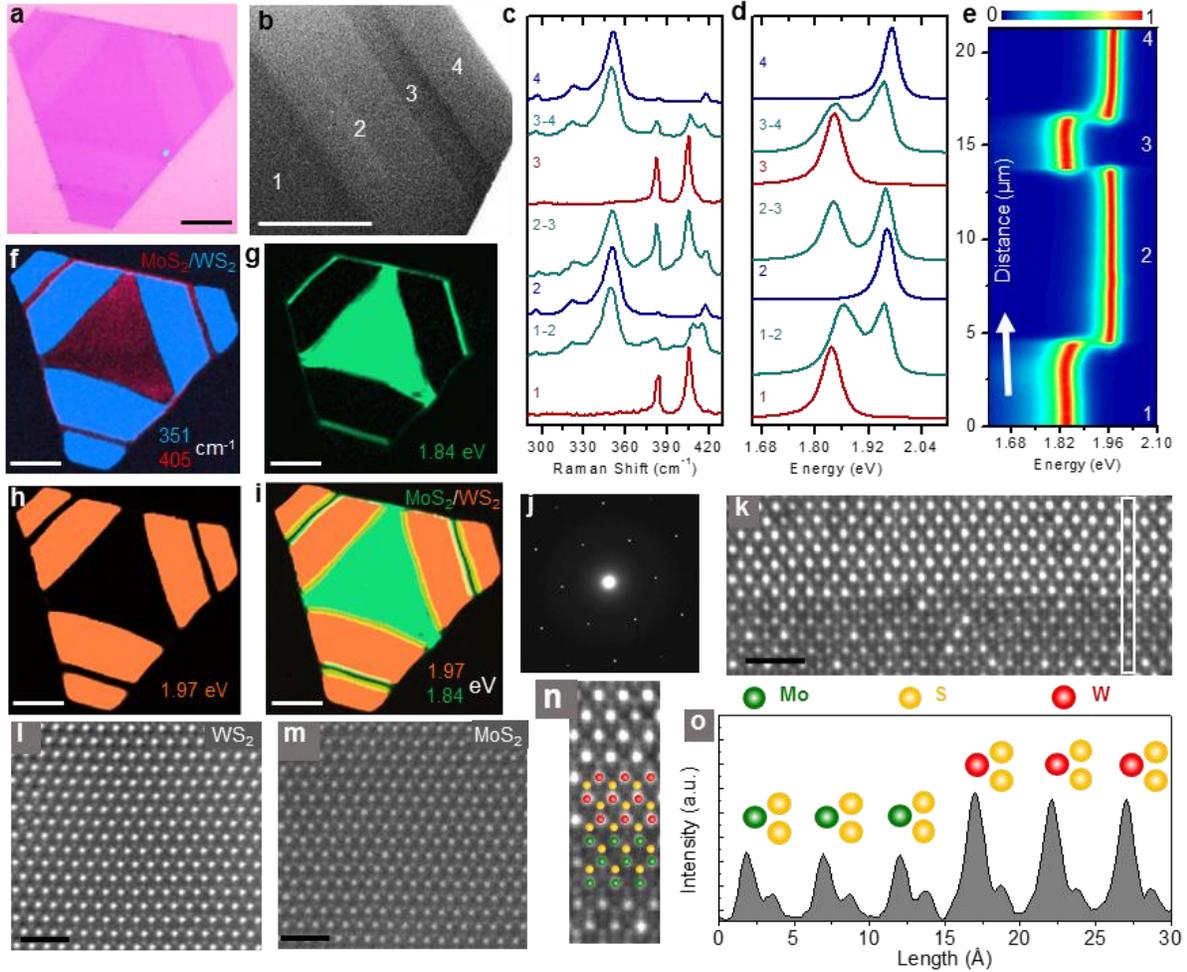

**Figure 3. Multi-junction lateral heterostructure based on $MoS_2$ and $WS_2$. a,** Optical image of a heterostructure composed of three $MoS_2$-$WS_2$-junctions. **b,** SEM image of the right-top corner of the heterostructure. **c, d,** Raman and PL spectra, respectively, at points 1, 2, 3, 4, in (**b**), as well as at the junctions 1-2, 2-3 and 3-4. **e,** Normalized PL color contour plot along a direction perpendicular to the interfaces, where the white arrow indicates the growth direction. **f,** Composite Raman intensity map . **g,** PL intensity map at 1.84 eV ($MoS_2$ domain). **h,** PL intensity map at 1.97 eV ($WS_2$ domain). **i,** Composite PL map. Scale bars (**a, b,** and **f-l**), 10 µm. **j,** Electron diffraction pattern of the heterostructure. **k, i, m,** Atomic resolution HAADF-STEM image of a $MoS_2$-$WS_2$ interface, pure $WS_2$ and $MoS_2$ regions, respectively. **n,** Atomic balls model superimposed on the Atomic resolution HAADF-STEM image of the interface. **o,** Electron intensity profile along the white box in (**k**). Scale bars (**k-l**), 1 nm.

Figure 3f shows the composite Raman intensity map of a three-junction triangular heterostructure, built by superimposing the 251 cm$^{-1}$ (*2LA(M)* of WS$_2$) and the 405 cm$^{-1}$ (*A$_{1g}$* of MoS$_2$) modes, which allows us to visualize the sequential lateral integration of MoS$_2$ and WS$_2$ domains (individual maps in Extended data Figs. E5d and E5e). The corresponding PL intensity maps at 1.84 eV (Fig. 3g) and at 1.97 eV (Fig. 3h) and the composite map (Fig. 3i) are also consistent with emissions from homogeneous monolayers of each material composing the heterostructure. The single-crystalline nature of the islands in the heterostructure was also verified by electron diffraction measurements (Fig. 3j). Z-contrast images from the inner regions of each domain (Figs. 4l and 4m), confirm the high-purity of both compounds. The high quality and the seamless nature of these interfaces produced by lateral epitaxy was also verified by Z-contrast STEM imaging (Figs. 3k to 3o).

To date, only single-junction heterostructures based on pure TMDs have been reported. However, it is important to develop procedures to make heterostructures containing TMD alloys with sharp band-gap discontinuities at their interfaces, this will expand our ability to engineer a wide range of 2D devices. With this objective, we extended the methodology described in the previous sections, to grow multi-junction heterostructures based on ternary alloys (MoS$_{2(1-x)}$Se$_{2x}$ - WS$_{2(1-x)}$Se$_{2x}$). To this effect, we used solid sources containing different metal and chalcogen atoms (e.g. MoSe$_2$ and WS$_2$ or MoS$_2$ and WSe$_2$). Prior studies indicate that the internal energy of mixing, for nearly all configurations of MoS$_{2(1-x)}$Se$_{2x}$ or WS$_{2(1-x)}$Se$_{2x}$ alloys. is negative and hence the alloys are energetically favored over the binary segregates.[24, 25]

Figures 4a and 4b show the optical image and the composite PL map, respectively, of a three-junction lateral heterostructure composed of monolayered alloys (MoS$_{2(1-x)}$Se$_{2x}$ - WS$_{2(1-x)}$Se$_{2x}$ - MoS$_{2(1-x)}$Se$_{2x}$ - WS$_{2(1-x)}$Se$_{2x}$), These were grown in a sequence of 12-0.3-3-0.7 min; *via* a similar carrier gas switching procedure using a mixture of MoS$_2$ and WSe$_2$ as the solid source. The sequential morphology observed in this case is close to conformal, indicating that the growth rate for each material in the heterostructure is isotropic (Extended data Fig. E6a). The morphology contrasts with the convex isosceles trapezoid observed for the pure sulfur-based heterostructures, and suggests that the Se/S ratio could affect the kinetics of the growth along the different crystallographic directions and determine the shape of the islands. The composite PL map shown in Fig. 4b for emissions at 1.61 and 1.71 eV across the entire triangular domain indicates a well-defined contrast between the MoS$_{2(1-x)}$Se$_{2x}$ and the WS$_{2(1-x)}$Se$_{2x}$ domains (Fig. 4a). Detailed PL spectra recorded from points 1 through 4 as well as from a line scan across the three-junctions is shown in Extended data Fig. E6.

By changing the solid sources to a mixture of MoSe$_2$ and WS$_2$, one can obtain a similar three-junction heterostructure, see Fig.4c, with different chalcogen (S/Se) compositions (Extended data Fig. E7). The corresponding composite PL map at 1.67 eV and at 1.8 eV is shown in Fig. 4d, and clearly reveals the presence of a sequence of alloyed domains in the heterostructure (detailed analysis in Extended data Fig. E8). The contour plot of the normalized PL intensity acquired along a line perpendicular to the interfaces is shown in Fig. 4e. The position of the PL peak is constant across each individual domain, suggesting homogeneity in their chemical composition. Additionally, there is a sharp discontinuity in the position of the PL peak at each interface. As discussed in the previous sections, this is a characteristic of abrupt junctions. TEM analysis indicates that the different domains are composed of ternary alloys of $MoS_{2(1-x)}Se_2$ or $WS_{2(1-x)}Se_2$. Figure 4f shows a Z-contrast image of a $WS_{2(1-x)}Se_{2x}$ domain. A line profile (Fig. 4g) reveals four distinct electron intensities associated with the metal sites (W in this case) and with three different configurations of the chalcogen sites (e.g. $S_2$, $Se_2$ or SSe).

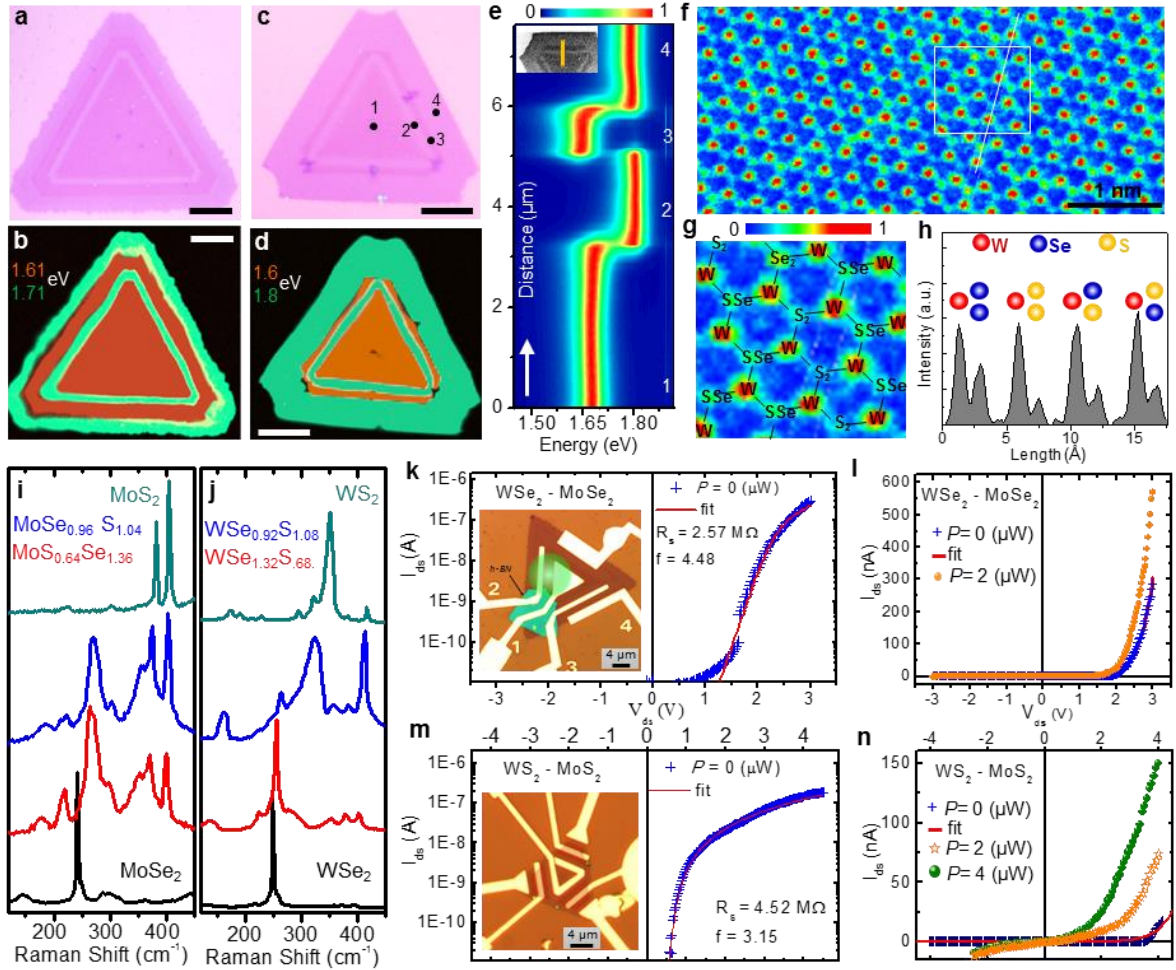

**Figure 4. Synthesis of three-junction lateral heterostructures based on $MoX_2$-$WX_2$ ($X_2=S_{2(1-n)}Se_{2n}$), and device characterization.** Optical image of a three-junction heterostructure, **a**, $MoS_{0.64}Se_{1.36}$-$WS_{.68}Se_{1.32}$ and **b**, the corresponding composite PL map at 1.61 eV and 1.71 eV. **c**, Optical image of a three-junction heterostructure composed of $MoS_{1.04}Se_{0.96}$ and $WS_{1.08}Se_{0.92}$. **d**, Corresponding composite PL map at 1.6 eV and 1.8 eV. Scale bars, 10 μm. **e,** Normalized PL color contour plot along a direction perpendicular to the interfaces (SEM image in the inset). **f,** Atomic resolution HAADF-STEM image of a $WS_{2(1-x)}Se_{2x}$ domain. **g**, Magnified image of the region enclosed by the box in (**f**) showing the different configurations of chalcogen sites. **h**, Electron intensity profile along the line in (**f**). **k,l,** Semi-logarithmic plot of the current as a function of the bias voltage ($I_{ds}$-$V_{ds}$) for $WSe_2$-$MoSe_2$ and $WS_2$-$MoS_2$ heterostructures, respectively. Traces collected under dark conditions indicate diode behavior. Fitting to the diode equation (red curves) with a series resistor yields ideality factors of $f = 4.47$ and $f = 3.15$, respectively. Insets: optical images of the devices. Scale bar, 4 µm. **m,n,** Photocurrent as a function of the bias voltage under several illumination powers under a wavelength λ = 532 nm having a spot size of 10 µm.

The concentration (*x*) in monolayer $MoS_{2(1-x)}Se_{2x}$ and $WS_{2(1-x)}Se_{2x}$ alloys can be calculated from the measured PL peak positions according to the Vegard's law,

$$E_g(MS_{2(1-x)}Se_{2x}) = (1-x)E_g(MS_2) + xE_g(MSe_2) - bx(1-x); \text{ where M= Mo, W}$$

Considering the band gap bowing parameters *b* = 0.05 and *b* = 0.04 for the Mo-based and W-based alloys, respectively,[24] the combination of compositions for the heterostructure in Fig. 4a is calculated to be $MoS_{0.64}Se_{1.36}$ (x=0.68) and $WS_{0.68}Se_{1.32}$ (x=0.66). Similarly, for the heterostructure shown in Fig. 4c, we obtained $MoS_{1.04}Se_{0.96}$ (x=0.48) and $WS_{1.08}Se_{0.92}$ (x=0.46). Noteworthy that a complete miscibility of (S, Se) was achieved for these individual $MoSe_{2(1-x)}S_{2x}$ and $WS_{2(1-x)}Se_{2x}$ domains. This is in agreement with the fact that the alloys tend to maximize the number of dissimilar atom pairs (S−Se) by minimizing the neighboring sites having (Se-Se) or (S-S) atoms in order to achieve the lowest energy configuration (Fig.4g).[24] The Raman spectra (Figs. 4i and 4j) collected from the different domains, as shown in Figs. 4d and 4e, also provides a clear indication for spectrally distinct alloy configurations (details in the supporting information and Extended Table1). This is the first demonstration of an alloy-based lateral heterostructure.

In order to investigate the electrical response of the $WX_2$-$MoX_2$ lateral heterojunctions, Ti/Au metallic contacts were separately deposited on both the $WX_2$ and the $MoX_2$ monolayers, (see Supplementary Materials for details). The insets in Figs. 4k and 4l show the optical images of the fabricated devices with their respective *I-V* characteristics plotted in a logarithmic scale. The *I-V* characteristic displays current rectification, demonstrating the presence of a *p-n* junction with threshold voltages of approximately 1.4 V and 0.87 V in forward bias configuration for $WSe_2$-$MoSe_2$ and $WS_2$-$MoS_2$, respectively. A reverse bias current of less than 1 pA and a rectification factor of $10^5$ proves promising characteristics for low-power electronics. The red curves are fits to the Shockley diode equation in the presence of a series resistor $R_s$:[26]

$$I_{ds} = \frac{nV_T}{R_s} W_0\left\{\frac{I_0 R_s}{fV_T} exp\left(\frac{V_{ds} + I_0 R_s}{fV_T}\right)\right\} - I_0$$

where $V_T$ is the thermal voltage at a temperature *T*, $I_0$ is the reverse bias current, *f* is the diode ideality factor (*f* = 1 for an ideal diode) and $W_0\{x\}$ is the Lambert function. The best fits, yield ideality factors of *f* = 4.47 and *f* = 3.15 with series resistances ranging from 2 MΩ to 5 MΩ for $WSe_2$-$MoSe_2$ and $WS_2$-$MoS_2$ junctions, respectively. Deviations from the ideal diode response could have different origins including recombination processes,[26] role of the Schottky barriers at the electrical contacts, or Fermi level pinning. The optoelectronic response of the heterostructures was further evaluated by exciting with a 532 nm laser. A considerable photocurrent is measured under laser illumination (Figs. 4m, n). The obtained photo-responsivities for $WSe_2$-$MoSe_2$ and $WS_2$-$MoS_2$ were 144 mAW$^{-1}$ ($V_{ds}$ = 3V) and 34 mAW$^{-1}$ ($V_{ds}$= 4V), respectively. The photoresponsivity was evaluated by subtracting the dark current. Further investigation is the scope of our future work.

A preliminary study was performed to evaluate the interaction mechanism between water vapor and the $MoX_2$ as well as $WX_2$ bulk powders. By allowing the solid precursor to interact with water vapor at 1060 °C for a prolonged time (>20-30 min), it was found that different Mo( or W) oxide phases evolve, which are assumed to be the main driving force behind the selective growth of the individual compounds. It can be clearly seen from the Raman spectra (Extended data Fig. E9a) that $MoO_2$ is the dominating phase evolving during the oxidation of $MoS_2$ or of $MoSe_2$. A previous report also confirmed that the main solid product during $MoS_2$ oxidation under water vapor at temperatures above 1000 °C is $MoO_2$[27] rather than $MoO_3$, which tends to be a stable phase under various reactive gas environments.[28] This oxidized Mo is in turn susceptible to volatilization in the presence of water vapor, being subsequently transported and saturating on the substrate at lower temperatures. The re-condensed $MoO_2$ vapor interacts with the already existing $H_2X$ - the oxidation by-products, to form $MoX_2$ at temperatures ranging between 650 and 800 °C.

Apparently, the overall oxidation reaction between $MoX_2$ and $H_2O$ at 1060 °C follows $MoX_2$ (s) + $nH_2O$ (g) → $MoO_2$ + $mH_2X$ (g) + $yH_2$ (g) + $zSO_2$ (g) (where X= S, Se). Subsequently, it is accompanied by a process of metal-oxide vaporization and re-condensation according to: $MoO_2$ + $2H_2X$ → $MoX_2$ + $2H_2O$ (where X= S, Se).[29, 30] This leads to the formation of $MoX_2$ domains. Interestingly, this reaction can be terminated abruptly by changing the carrier gas from $N_2+H_2O$ to $Ar+H_2$, which reduces (*via* hydrogen) $MoO_2$ into Mo(s) + $MoO_2$(s) at the surface of the source. This suddenly depletes the source of $MoO_2$ volatized vapors thus ending the growth.

In contrast, $WX_2$ has a different oxidative behavior under the above conditions, where different $W_xO_y$ oxide phases are observed in the Raman Spectra (Extended data Fig. E9b and Supporting Information Table 2). The oxidation of W is dominated by either $W_{20}O_{58}$ (in $N_2+H_2O$ vapor) or by $W_2O$ + $WO_x$ in the presence of a reducing gas ($Ar+H_2$ with $H_2O$ vapor). The oxidation reaction at 1060 °C (at the source) will depend on the gas environment as follows:

1) $N_2$ + $H_2O$ vapor: $WX_2$ + $nH_2O$ → $W_xO_y$ (s) + $WO_x(OH)_2$ (g) + $mH_2X$ (g) + $yH_2$ (g) + $zXO_2$ (g)

2) $H_2$ + $H_2O$ vapor: $WX_2$ + $nH_2O$ → W(s) + $WO_2$ (s,g) + $WO_x$ (s,g) + $mH_2X$ (g) + $yH_2$ (g) + $zXO_2$ (g)

From the above reactions, volatile hydroxides are expected to form under $N_2+H_2O$ vapor environment. However, hydroxide species mostly condense below 500 °C.[31] Hence, tungsten deposition on the substrate will not occur for temperatures exceeding 600 °C. In contrast, in the presence of a reducing agent ($H_2$), re-condensation of $WO_x$ gases and subsequent chalcogenization is expected to occur on the surface of the substrate following the reaction: $WO_x$ + $2H_2X$ → $WX_2$ + $2H_2O$.

The appearance of different Mo and W oxidation states can also be directly observed from the changes in color suffered by the solid precursors after exposition to different gaseous environments; $MoO_2$ (brown), $W_{20}O_{59}$ (blue-violet), $WO_2$ (chocolate brown) (see Supporting Information and Extended data Fig. E9 c-j).

The growth mechanism can be summarized as follows. The selective growth of MoX$_2$ or WX$_2$ monolayers can be achieved simply by controlling the carrier gas environment. N$_2$+ H$_2$O vapor (without H$_2$) favors evaporation of both Mo and W precursors, but only Mo precursors are deposited on the substrate. A sudden switch of the carrier gas to (Ar+H$_2$), quickly depletes the supply of Mo precursors, while continuing to supply W precursors due to the slower reduction rate of WO$_x$. This vapor-phase modulation of the oxide species is the key driving force for the observed sequential growth of lateral heterostructures. A more detailed chemical analysis including the type of gaseous by-products in conjunction with theoretical models are ongoing.

The developed method is unique, versatile and scalable; allowing to control the sequential edge-epitaxy of different transition metal dichalcogenides. The multi-junction lateral 2D heterostructures of binary as well as alloy-based materials, with tuned optical properties, covered three different regions of the spectrum (1.52-1.6 eV, 1.6-1.8 eV and 1.84-1.97 eV). The ability to tune the band gap in heterostructures opens up a wide range of possibilities for designing spectral-selective complex 2D materials for optoelectronic applications.

**Acknowledgements** This work was supported by the National Science Foundation (DMR-1557434). L.B acknowledges the U.S. Army Research Office MURI Grant W911NF-11-1-0362 (synthesis and physical characterization of two-dimensional materials and their heterostructures) and the Office Naval Research DURIP Grant# 11997003 (stacking under inert conditions). TEM work was performed at the NHMFL which is supported by the NSF Cooperative Agreement No. DMR-1157490 and the State of Florida. P.K.S. and H.R.G. thank Prof. Monica A. Cotta for comments.

**Author Contributions** H.R.G. supervised the project. P.K.S. and H.R.G. conceived the idea and designed the experiments. P.K.S. performed the synthesis, Raman and Photoluminescence characterizations, and related analysis. S.H. and L.B. perform device fabrication, electrical measurements and analysis. Y.X. conducted aberration-corrected STEM imaging with assistance from P.K.S. and H.R.G. H.R.G carried out TEM data analysis. P.K.S and H.R.G. analyzed the results and wrote the paper with inputs from L.B., S.M. and Y.X. All authors discussed the results and commented on the manuscript.

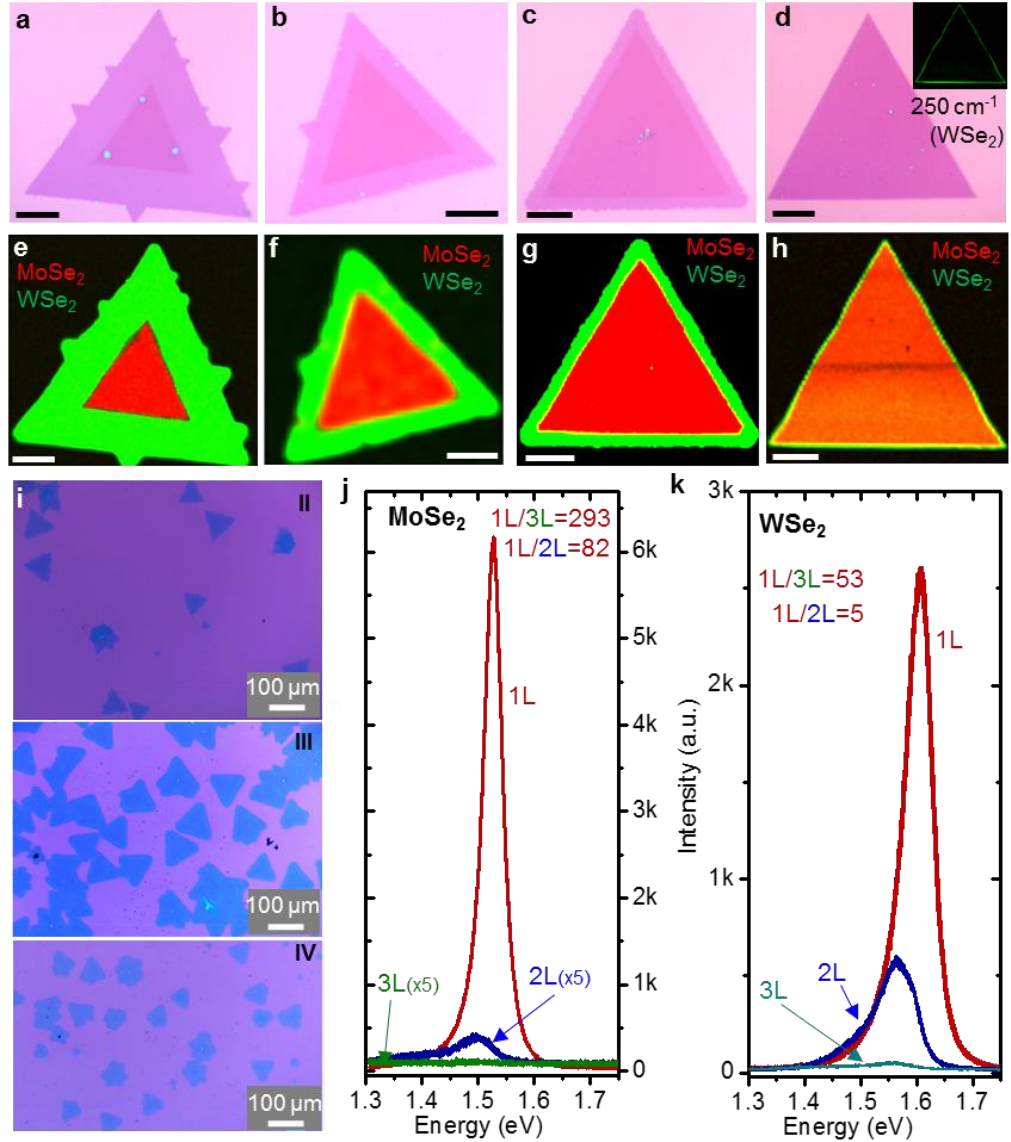

**Extended data Figure E1. Growth of single-junction MoSe$_2$ - WSe$_2$ lateral heterostructures. a,** and **d,** Optical images of single-junction MoSe$_2$-WSe$_2$ monolayer lateral heterostructure with different WSe$_2$ lateral growth time of 80, 45, 30, 15 sec respectively. **e-g,** Composite PL map corresponding to optical images (a-c), respectively, at 1.53 eV (MoSe$_2$ domain) and 1.6 eV (WSe$_2$ domain). **h,** Composite Raman map corresponding to optical image (d), at frequency 240 cm$^{-1}$ (MoSe$_2$ domain) and 250 cm$^{-1}$ (WSe$_2$ domain); in-set in (d), shows the Raman map of narrow WSe$_2$ shell which is difficult to visualize in the optical image. **i,** Low magnification optical images of MoSe$_2$-WSe$_2$ single-junction heterostructure, as shown in (b), obtained at different distance from the source precursor as mentioned in the Figure **1l-m** (region II to IV). PL spectra of monolayer (1L), bilayer (2L) and tri-layer (3L) corresponding to **j,** MoSe$_2$ and **k,** WSe$_2$ domain. Scale bars (**a-h**), 10 μm.

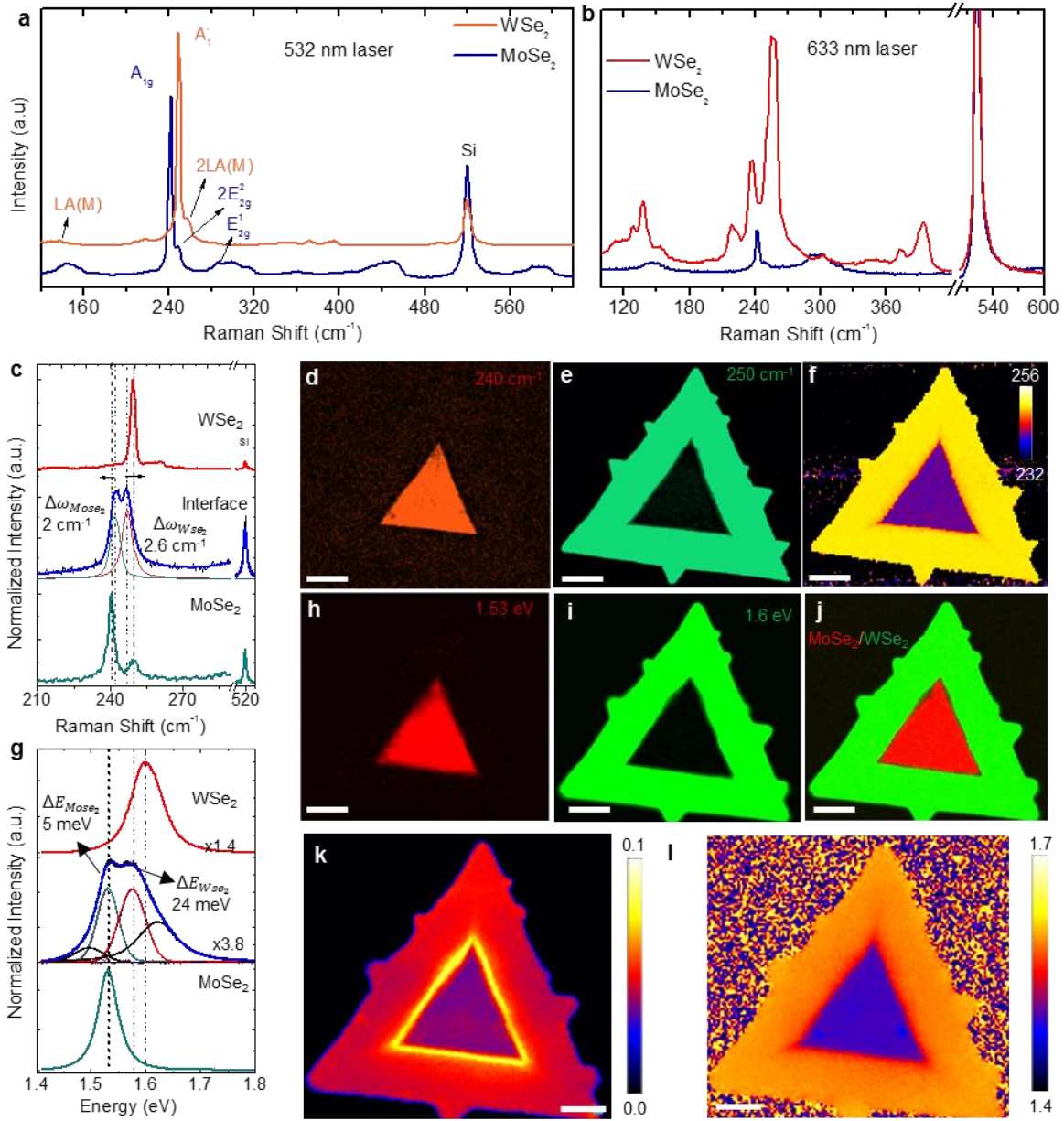

**Extended data Figure E2. Optical properties of single-junction MoSe₂-WSe₂ lateral heterostructure.** Raman spectra of MoSe₂ and WSe₂ domains from a single-junction MoSe₂-WSe₂ monolayer lateral heterostructure using **a**, 514 nm; **b**, 613 nm laser excitation; and **c**, showing the Raman spectra at the interfaces. Composite Raman intensity map **d,** at frequency 240 cm⁻¹ (MoSe₂ domain); **e**, 250 cm⁻¹ (WSe₂ domain) and **f**, position map corresponding to optical image in Fig. **1a**. **g**, PL spectra of MoSe₂, WSe₂ domains and at the interface of the MoSe₂-WSe₂ single-junction monolayer lateral heterostructure as shown in Fig. **1a**. PL intensity map **h,** at 1.53 eV (MoSe₂ domain); **i**, 1.6 eV (WSe₂ domain); **k**, peak width (in eV) map and **l**, position map corresponding to optical image in Figure **1a**. Scale bars (**d-l**), 10 μm.

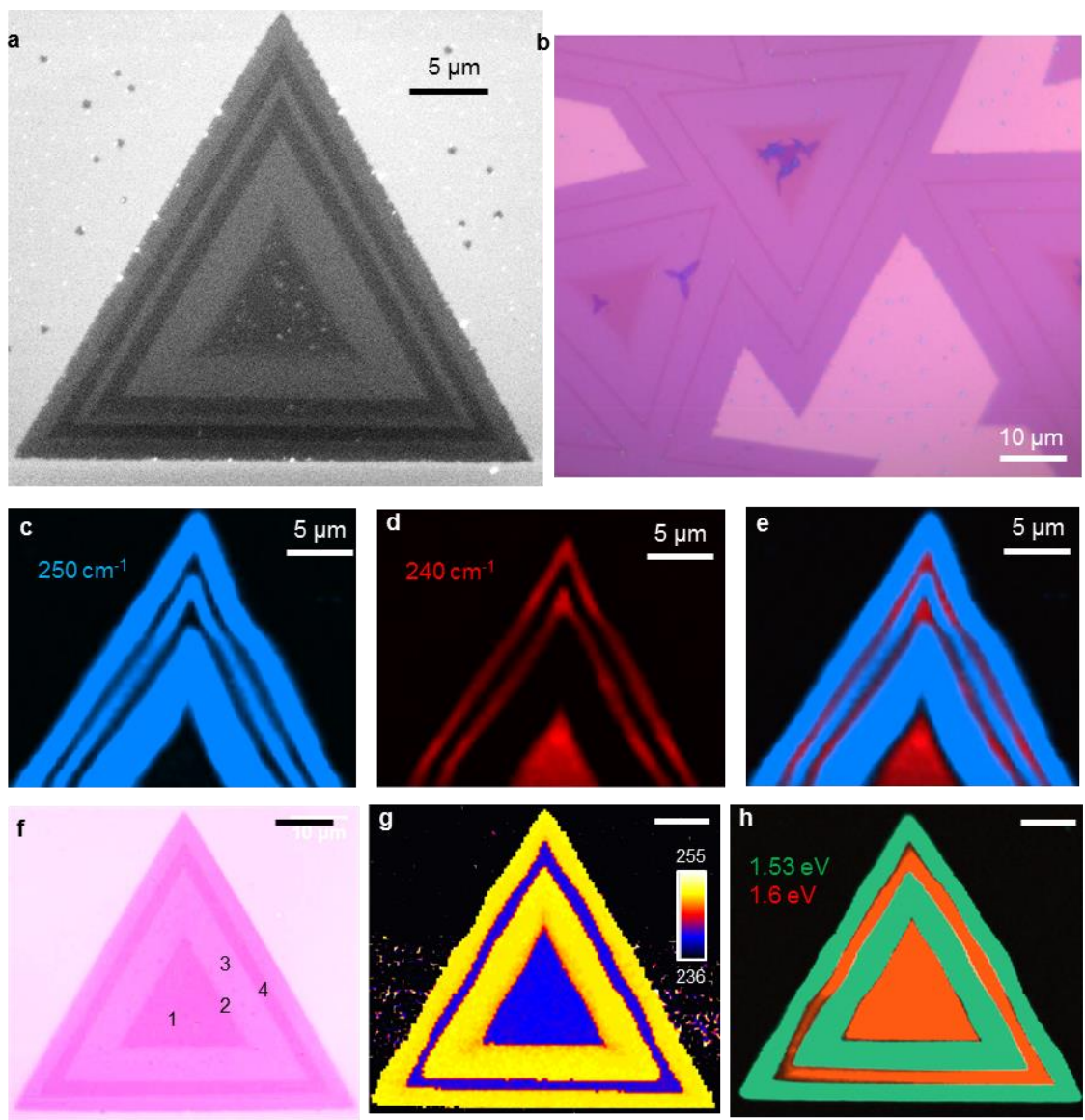

**Extended data Figure E3. Optical properties of multi-junction MoSe$_2$-WSe$_2$ lateral heterostructure. a**, Typical scanning electron microscope image of a five-junction MoSe$_2$-WSe$_2$ monolayer lateral heterostructure corresponding to Fig. **1e**. **b,** Optical microscope image of large area view of five-junction MoSe$_2$-WSe$_2$ lateral heterostructure corresponds to Fig. **1f** showing the conformal growth of respective MoSe$_2$ or WSe$_2$ domains. Raman intensity map of five-junction MoSe$_2$-WSe$_2$ lateral heterostructure corresponding to Fig. **1e** and **j**, at frequencies **c**, 250 cm$^{-1}$ (WSe$_2$ domain); **d**, 240 cm$^{-1}$ (MoSe$_2$ domain), and **e**, composite Raman map image at 250 and 240 cm$^{-1}$. **f**, Optical image of a three-junction MoSe$_2$-WSe$_2$ monolayer lateral heterostructure corresponding to Fig. **2c** (in-set). **g,** Raman peak position mapping between 236-255 cm$^{-1}$ spectral ranges; and **h**, composite PL intensity mapping at 1.53 eV (MoSe$_2$ domain) and 1.6 eV (WSe$_2$ domain). Scale bars (**f-h**), 10 μm.

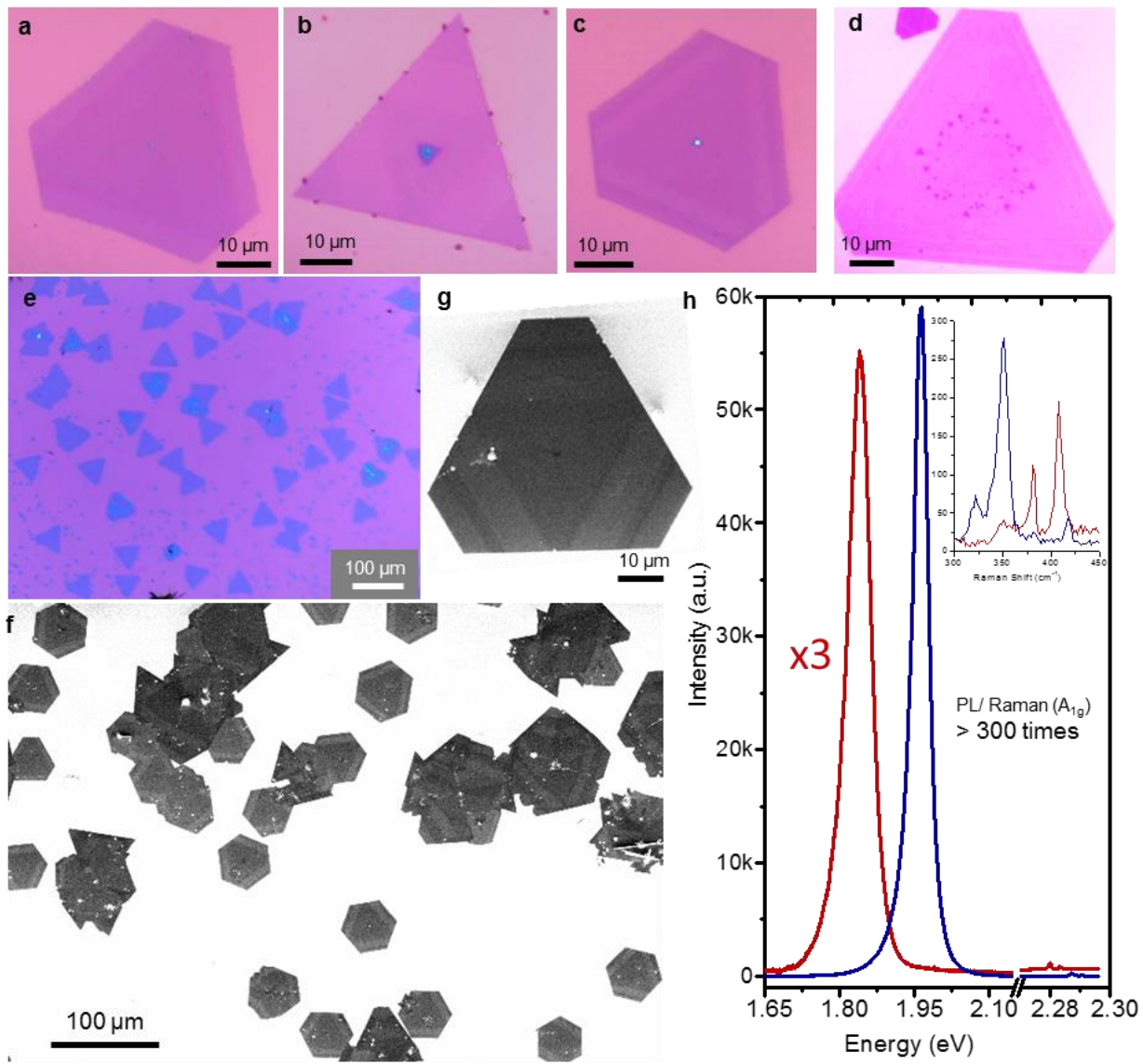

**Extended data Figure E4. Growth of multi-junction MoS$_2$-WS$_2$ lateral heterostructure. a-d,** Optical images of MoS$_2$-WS$_2$ monolayer lateral heterostructure of different; **a**, single-junction, **b**, two-junction, **c**, three-junction, **d**, five-junction and **e**, typical low magnification optical image corresponding to (**d**). **f**, SEM images of three-junction MoS$_2$-WS$_2$ lateral heterostructures, (corresponding to optical image in (**c**)) and **g**, SEM image a three-junction single island (Fig. **3b**). **h**, Typical PL spectra from MoS$_2$ and WS$_2$ domains corresponding to the three-junction heterostructure in (**g**). The strong PL/Raman (A$_{1g}$) (> 300 times) indicates the monolayer nature as well as high optical quality of the as grown heterostructure.

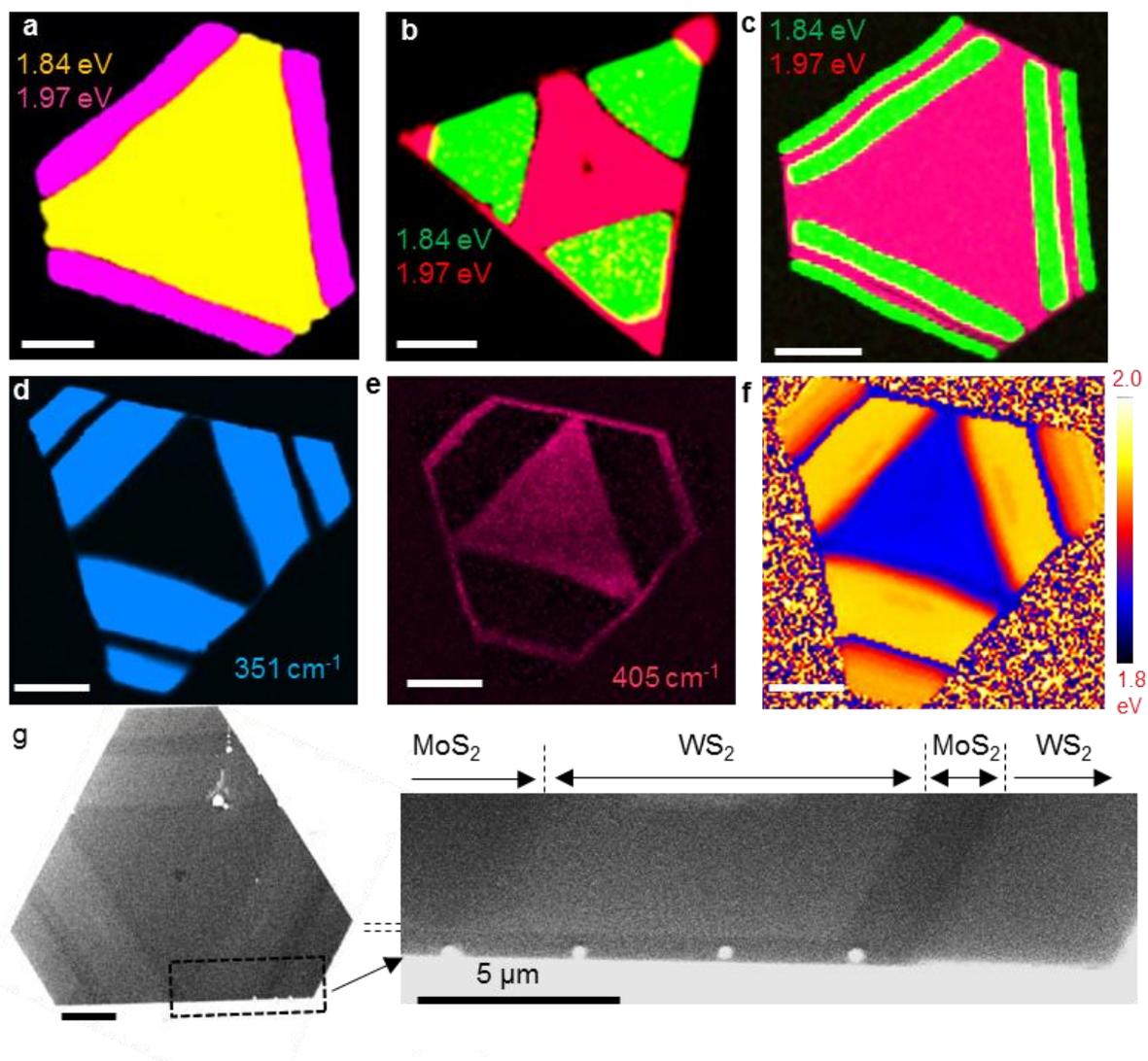

**Extended data Figure E5. Optical properties of MoS$_2$-WS$_2$ lateral heterostructures.** Composite PL intensity mapping of **a**, single-junction; **b**, two-junction; **c**, three-junction MoS$_2$-WS$_2$ monolayer lateral heterostructure corresponding to optical images in Extended data Figure **E4a-c**, respectively, at 1.84 eV (MoS$_2$ domain); **i**, 1.97 eV (WS$_2$ domain). Raman intensity mapping **d,** at frequency 351 cm$^{-1}$ (WS$_2$ domain); **e**, 405 cm$^{-1}$ (MoS$_2$ domain), and **f**, PL position mapping corresponding to optical image in Fig. 3**a**. **g**, SEM image a three-junction MoS$_2$-WS$_2$ monolayer lateral heterostructure island with high magnification image (box region, right panel) showing the lateral connectivity between respective domains of MoS$_2$ or WS$_2$. Scale bars (**a-g**), 10 μm.

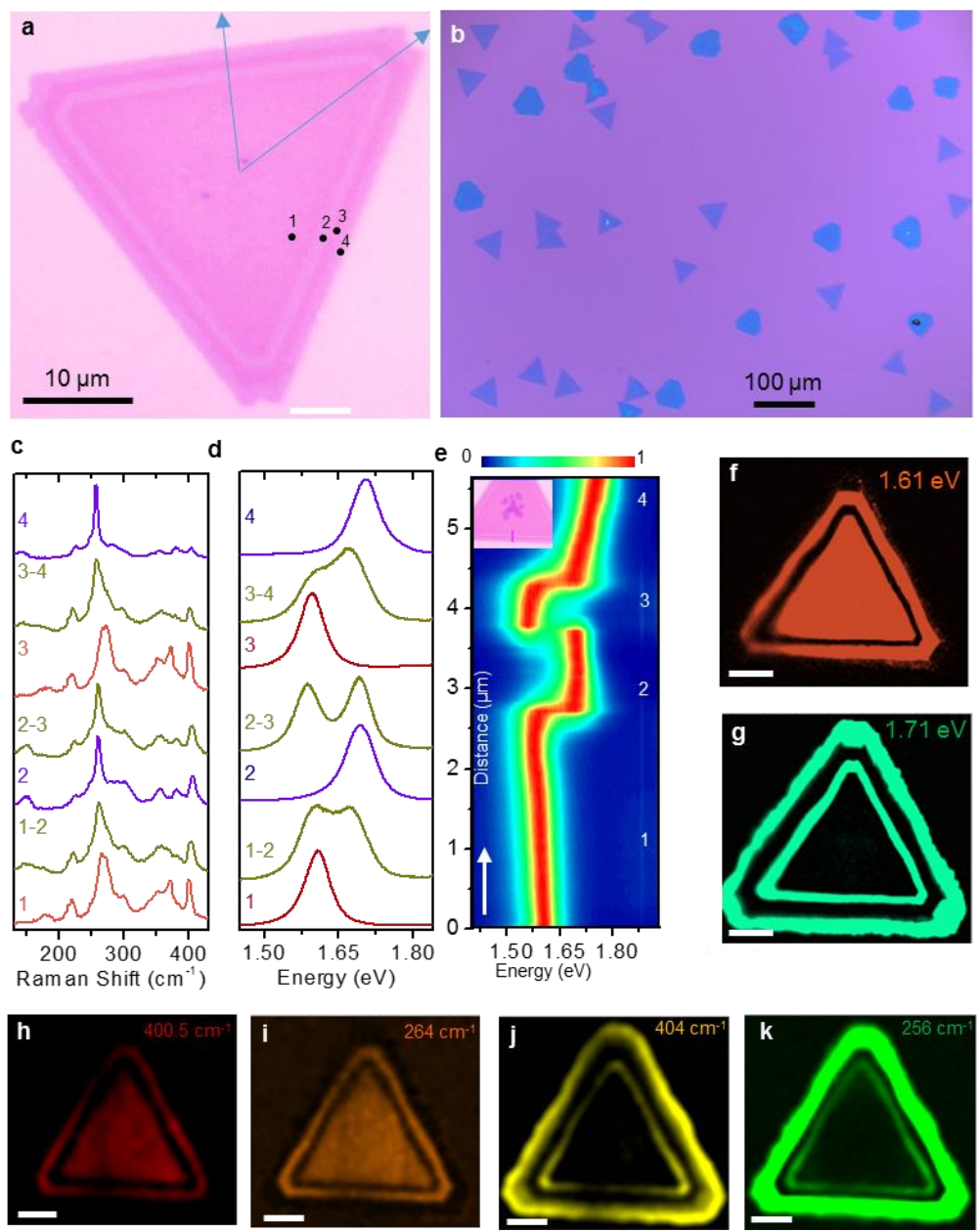

**Extended data Figure E6. Growth of three-junction MoS$_{0.64}$Se$_{1.36}$ - WSe$_{1.32}$S$_{.68}$ lateral alloy heterostructure. a**, Optical image of a three-junction MoS$_{2(1-x)}$Se$_{2x}$ - WSe$_{2x}$S$_{2(1-x)}$ monolayer lateral heterostructure, and **b**, the corresponding low magnification optical image. **c, d,** Raman and PL spectra, respectively, corresponding to optical image in (a) and between points 1 through 4; and **e**, normalized PL contour color plot along the direction perpendicular to the interfaces as shown in the in-set optical image. The PL intensity maps at **f**, 1.61 eV (MoS$_{0.64}$Se$_{1.36}$ domain) and **h**, 1.71 eV (WSe$_{1.32}$S$_{.68}$ domain) corresponding to optical image in Fig. **4a**. Raman intensity maps (Fig. **4a)** at frequency **c**, 400.5 cm$^{-1}$ (A$_{1g(S-Mo)}$ modes); **d**, 264 cm$^{-1}$ (A$_{1g(Se-Mo)}$ modes); **e**, 404 cm$^{-1}$ (A$_{1g(S-W)}$ mode); and **f**, 256 cm$^{-1}$ (A$_{1g(Se-W)}$ mode). Scale bars (**e-j**), 10 μm.

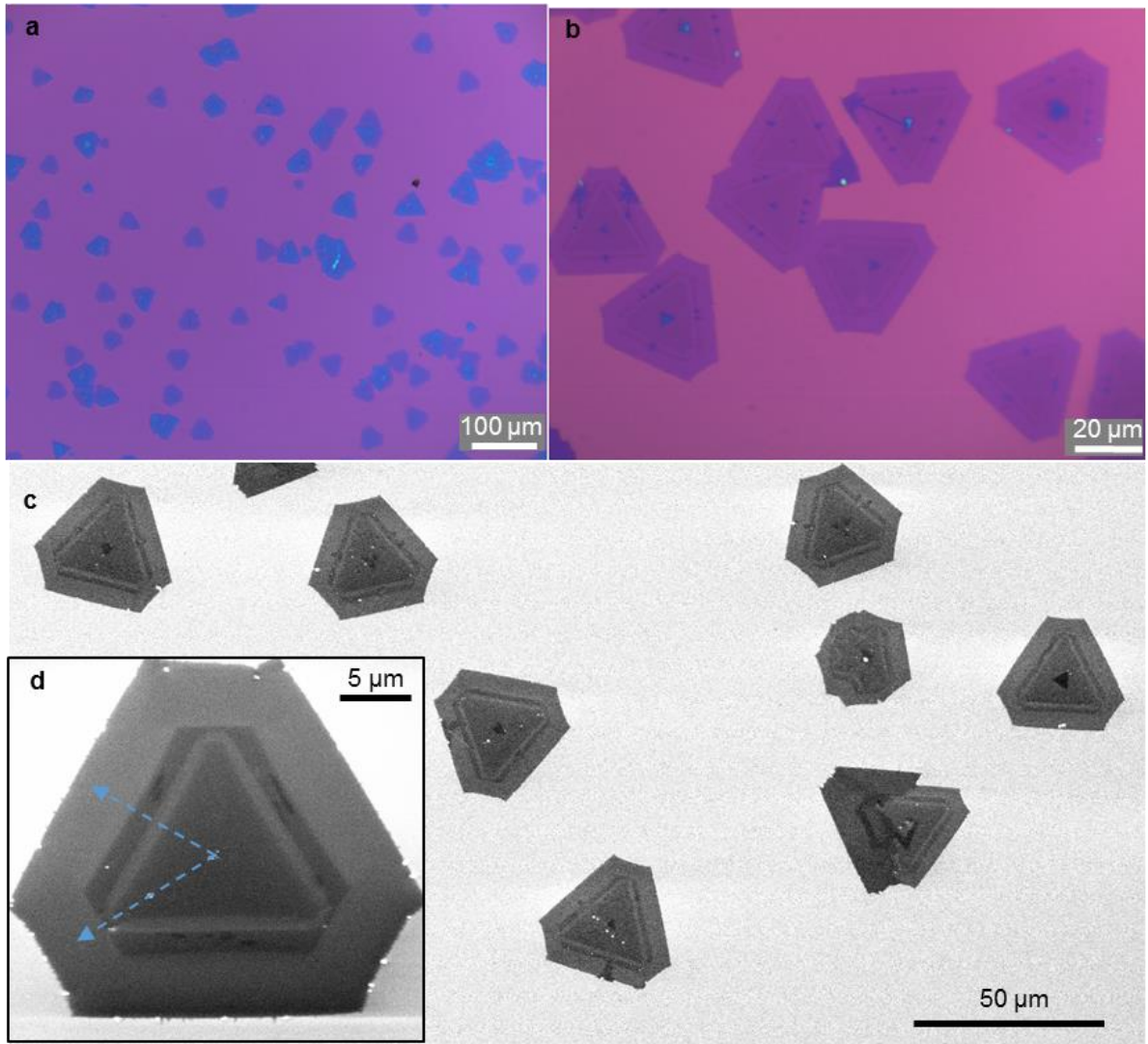

**Extended data Figure E7. Growth of three-junction MoSe$_{0.96}$S$_{1.04}$-WSe$_{0.92}$S$_{1.08}$ lateral alloy heterostructure. a-b**, Low magnification optical image of three-junction MoS$_{2(1-x)}$S$_{2x}$ - WS$_{2(1-x)}$Se$_{2x}$ monolayer lateral heterostructure (corresponding to optical image in Fig. **4c**), and **c,** typical large area SEM image, and d, high magnification SEM image of a single island showing the presence of different growth rate along vertex and axial direction. The MoS$_{2(1-y)}$S$_{2y}$ growth along the vortex direction is less than that of axial direction.

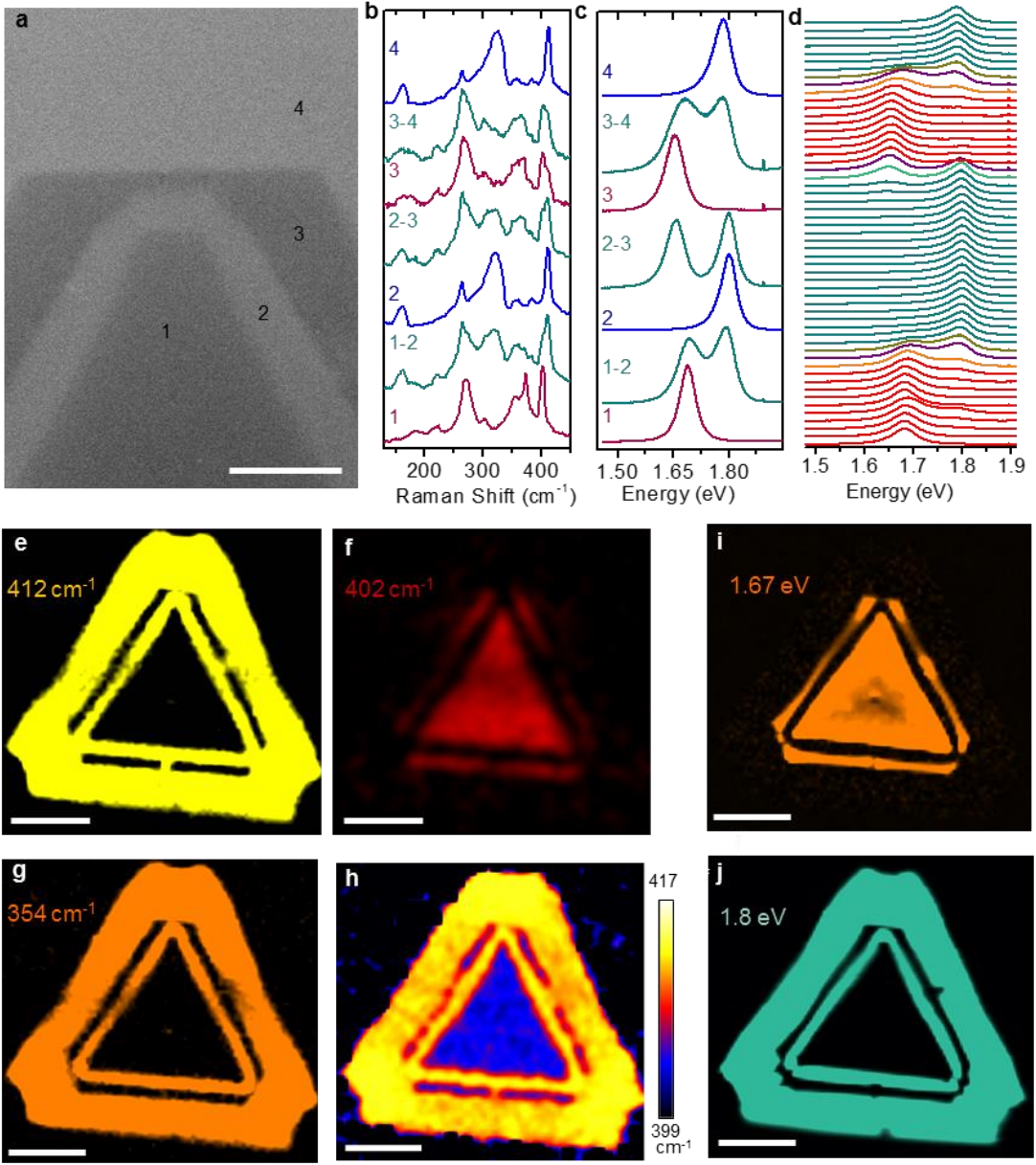

**Extended data Figure E8. Optical properties of multi-junction MoSe$_{0.96}$S$_{1.04}$-WSe$_{0.92}$S$_{1.08}$ lateral heterostructure. a**, SEM image of a three-junction MoS$_{2(1-x)}$S$_{2x}$ - WS$_{2(1-x)}$Se$_{2x}$ monolayer lateral heterostructure. Scale bar, 2 µm. **b, c**, Raman and PL spectra, respectively, between points 1 through 4; and interfaces. **d**, Normalized PL spectra from a line scan along three-junction as indicated in the inset in Fig. **4e.** Raman intensity map corresponding to optical image in Fig. **4c**, at frequency **e**, 412 cm$^{-1}$ (A$_{1g(S-W)}$ modes); **f**, 402 cm$^{-1}$ (A$_{1g(S-Mo)}$ mode), **g**, 354 cm$^{-1}$ (E$_{2g(S-W)}$ modes); and **h**, Raman position mapping between 399-417 cm$^{-1}$. There exist a thin line of MoS$_{2(1-y)}$S$_{2y}$ between WS$_{2(1-y)}$Se$_{2y}$ strip along the vertex direction which could not be resolved during the Raman Mapping. PL intensity map, corresponding to optical image in Figure **4c**, at **i**, 1.67 eV (MoSe$_{0.96}$S$_{1.04}$ domain) and **j**, 1.8 eV (WSe$_{0.92}$S$_{1.08}$ domain). Scale bars (**e-j**), 10 µm.

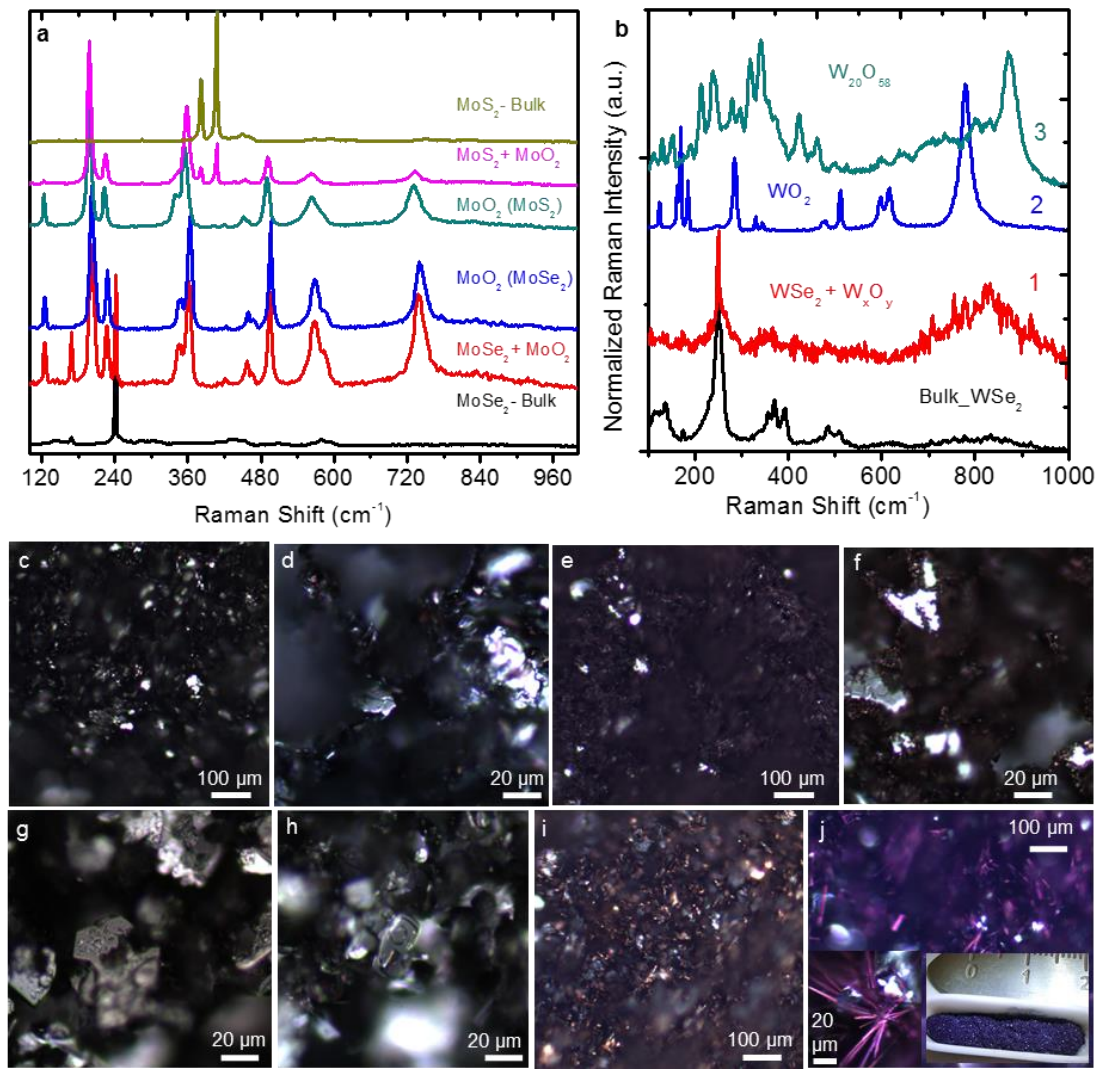

**Extended Figure E9. Effect of Water Vapor on the Solid Sources (MoX$_2$ and WX$_2$). a**, Raman spectral evolution of MoO$_2$ oxide phase from both MoSe$_2$ and MoS$_2$ solid sources upon reaction with a constant flow of N$_2$+ H$_2$O vapor for more than 20 min at 1060 °C (Supporting information Table 2). A direct visualization can be gained from the observed change in color of the source precursor at different condition; **c**, bulk powder of MoSe$_2$; **d**, Ar+H$_2$ (5%) through H$_2$O (200 sccm); **e**, N$_2$ through H$_2$O (200 sccm)- the chocolate brown refers to MoO$_2$ phase, and **f**, shining surface indicating the presence of metal Mo reduced from MoX$_2$ along with MoO$_2$ phase. **b**, Raman spectral evolution of different oxide phases of WX$_2$ (optical image in **g**) upon reaction with different reactive gas environment for more than 20 min at 1060 °C as follows. Case 1) Only Ar+H$_2$ (5%) through H$_2$O (200 sccm): the Raman spectra is composed of WSe$_2$, most likely a Se deficient surface as well as a mixture of complex oxide phases as indicated by the broad peak around 800 cm$^{-1}$, the optical image in (h). Case 2) First partially oxidized by N$_2$ + H$_2$O (5 min) followed by Ar+H$_2$ (5%) through H$_2$O (200 sccm) for 10 min. The dominating phase observed in the Raman spectra is WO$_2$. Case 3) Completely oxidized by N$_2$ + H$_2$O flow for 20 min. The dominating phase observed in the Raman spectra is W$_{20}$O$_{58}$. The optical images further confirmed the presence of different oxide phases in the source precursor; **i**, chocolate brown – WO$_2$ phase, and **j**, blue-violet indicating the presence of W$_{20}$O$_{58}$ phase; inset in **j**, the high magnification image (left panel) and the materials in alumina boat (Detailed in Supporting information Table 2).

**Table 1.** Raman and PL peak position of different heterostructure obtained in this study.

| Type of Heterostructure | 1st (core) | | 1st - 2nd Interface | | 2nd Layer | | 2nd - 3rd Interface | | 3rd Layer | | 3rd - 4th Interface | | 4th (Shell) | |
|---|---|---|---|---|---|---|---|---|---|---|---|---|---|---|
| | Raman cm$^{-1}$ | PL eV | Raman cm$^{-1}$ | PL eV | Raman cm$^{-1}$ | PL eV | Raman cm$^{-1}$ | PL eV | Raman cm$^{-1}$ | PL eV | Raman cm$^{-1}$ | PL eV | Raman cm$^{-1}$ | PL eV |
| MoSe$_2$/WSe$_2$<br><br>Ref. [6-8] | *145*<br>**240.5** *[A$_{1g}$]*<br>*248.5*<br>**286** *[E$^1{}_{2g}$]*<br>*361*<br>*450* | 1.52 | 243<br>246.4 | 1.535<br>1.575 | 249<br>259 | 1.59 | 241<br>249 | 1.525<br>1.59 | **241** | 1.52 | 241<br>248 | 1.53<br>1.58 | **249-250** *[A$_{1g}$]*<br>259<br>*119*<br>*128*<br>*138*<br>*358*<br>*358*<br>*372*<br>*395* | 1.60 |
| MoS$_2$/WS$_2$<br><br>Ref. [8-11] | **384** *[E$_{2g}$]*<br>**405** *[A$_{1g}$]* | 1.84 | 174<br>196<br>322<br>349.5<br>357<br>381<br>308.5<br>415.5 | 1.865<br>1.958 | 175<br>192<br>230<br>296<br>323<br>351<br>355<br>384<br>418 | 1.963 | 174<br>191<br>229<br>295<br>322<br>350<br>356<br>382<br>405<br>417.5 | 1.84<br>1.96 | **382**<br>**405** | 1.84 | 174<br>190<br>230<br>295<br>321.5<br>350<br>356<br>382<br>407<br>417 | 1.85<br>1.96 | 175 *[LA(m)]*<br>192<br>230 *[A$_{1g}$(m)-LA(m)]*<br>297 *[2LA(m)-2E$^2{}_{2g}$(m)]*<br>323<br>**351** *[2LA(m)]*<br>355 *[E$_{2g}$]*<br>384<br>**418** *[A$_{1g}$]* | 1.97 |
| MoS$_{2(1-x)}$Se$_{2x}$-WS$_{2(1-x)}$Se$_{2x}$<br><br>[MoSe$_{0.96}$S$_{1.04}$ WSe$_{0.92}$S$_{1.0}$]<br><br>Ref. [2-5] | 223<br>267<br>278<br>303<br>356<br>374<br>402.5 | 1.675 | 261.7<br>221<br>263<br>274<br>320<br>358<br>370<br>383<br>404<br>411 | 1.695<br>1.79 | 160<br>263<br>290<br>322<br>356<br>383<br>411.5 | 1.8 | 260<br>222<br>263<br>272<br>320.6<br>360<br>381<br>402.5<br>411.5 | 1.65<br>1.8 | 223<br>267<br>278<br>302.5<br>360<br>371<br>402.5 | 1.66 | 160<br>220<br>263<br>276<br>302.5<br>319<br>355<br>365<br>383<br>402.5<br>410 | 1.69<br>1.79 | 161.7<br>263<br>290<br>324<br>358<br>383<br>413 | 1.79 |
| MoS$_{2(1-x)}$Se$_{2x}$-WS$_{2(1-x)}$Se$_{2x}$<br><br>[MoS$_{0.64}$Se$_{1.36}$ WSe$_{1.32}$S$_{.68}$]<br><br>Ref. [2-5] | 178<br>219<br>264<br>299<br>352<br>370<br>400.5 | 1.61 | 138<br>221<br>261<br>277<br>300<br>357<br>370<br>381<br>404 | 1.61<br>1.68 | 138<br>225<br>259<br>291<br>354.5<br>379<br>406 | 1.69 | 148<br>223<br>259<br>263<br>279<br>299<br>356<br>370<br>381<br>404 | 1.595<br>1.685 | 176<br>219.5<br>268<br>300<br>352.5<br>372<br>400.5 | 1.6 | 221<br>257.5<br>263<br>273<br>299<br>356<br>379<br>402 | 1.6<br>1.68 | 141<br>225 A$_{1g}$(m)-LA$_{(S-W)}$<br>256 A$_{1g}$(Se-W)<br>284<br>354 E$_{2g}$(S-W)<br>379 A$_{1g}$(S-W-Se)<br>404 A$_{1g}$(S-W) | 1.715 |

**Table 2.** Major Raman peak positions of different solid sources, and their associated oxide phases. Bold numbers indicating the strong peak.

| MoSe$_2$ | MoS$_2$ | MoO$_2$ [12] | WSe$_2$ | WO$_2$ [13, 15, 16] | W$_{20}$O$_{58}$ [14] |
|---|---|---|---|---|---|
|  |  | 126 | 135 | 124 | 112 |
|  |  | **203** |  | 161 | 130 |
| **241** |  | 228 | **251** | 168.5 | 152 |
|  |  | 347 |  | 185 | 190 |
|  |  | **363** | 355 | 284 | **212** |
|  | **382** | 458 | 370 | 329 | **237** |
|  | **407.5** | **496** | 393 | 343 | 279 |
|  |  | 568 |  | 478 | 297 |
|  |  | **742** |  | 511 | **317** |
|  |  |  |  | 590 | **340** |
|  |  |  |  | 616 | 422 |
|  |  |  |  | 778 | 460 |
|  |  |  |  | 822 | 800 |
|  |  |  |  |  | 831 |
|  |  |  |  |  | **872** |

I. References

1. Lee, C. H. et al. Atomically thin p-n junctions with van der Waals heterointerfaces. Nature Nanotechnology 9, 676-681 (2014).
2. Jadczak, J. et al. Composition dependent lattice dynamics in MoS$_x$Se$_{(2-x)}$ alloys. Journal of Applied Physics 116 (2014).
3. Feng, Q. L. et al. Growth of Large-Area 2D MoS$_{2(l-x)}$Se$_{2x}$, Semiconductor. Advanced Materials 26, 2648-2653 (2014).
4. Feng, Q. L. et al. Growth of MoS$_{2(1-x)}$Se$_{2x}$ (x=0.41-1.00) Monolayer Alloys with Controlled Morphology by Physical Vapor Deposition. Acs Nano 9, 7450-7455 (2015).
5. Duan, X. D. et al. Synthesis of WS$_{2x}$Se$_{2-2x}$ Alloy Nanosheets with Composition-Tunable Electronic Properties. Nano Letters 16, 264-269 (2016).
6. Tonndorf, P. et al. Photoluminescence emission and Raman response of monolayer MoS$_2$, MoSe$_2$, and WSe$_2$. Optics Express 21, 4908-4916 (2013).